\documentclass[twocolumn,showpacs,preprintnumbers,amsmath,amssymb,aps,a4paper]{revtex4}

 \usepackage{graphicx}
 \usepackage{dcolumn}
 \usepackage{bm}
 \usepackage{amssymb,amsmath}
 \usepackage{verbatim}
 \usepackage{subfigure}
 \usepackage{textcomp}
 \newcommand{\YSO}{Y$_2$SiO$_5$ }

 \newcommand{\ket}[1]{\left|#1\right\rangle}

\begin{document}

  \title{Strong coupling cavity QED using rare earth metal ion dopants in
    monolithic resonators: what you can do with a weak oscillator.}

 \author{D. L. McAuslan}
 \author{J. J. Longdell}
 \email{jevon@physics.otago.ac.nz}
 \affiliation{Jack Dodd Centre for Photonics and Ultra-Cold Atoms, Department of Physics,
University of Otago, Dunedin, New Zealand.}
 \author{M. J. Sellars}
 \affiliation{Laser Physics Centre, Research School of Physical
    Sciences and Engineering, Australian National University,
    Canberra, ACT 0200, Australia.}
  \date{\today}
\begin{abstract}
  We investigate the possibility of achieving the strong coupling regime of
  cavity quantum electrodynamics using rare earth ions as impurities
  in monolithic optical resonators. We conclude that due to the weak
  oscillator strengths of the rare earths, it may be
  possible but difficult, to reach the regime where the single
  photon Rabi frequency is large compared to both the cavity and atom
  decay rates. However reaching the regime where the saturation photon
  and atom numbers are less than one should be much more achievable. We show that
  in this `bad cavity' regime, transfer of quantum states and an
  optical phase shift conditional on the state of the atom is still
  possible, and suggest a method for coherent detection of single dopants.
\end{abstract}

  \pacs{3.67.Lx,82.53.Kp,78.90.+t}
  \keywords{Quantum state tomography, Quantum computation, Coherent Spectroscopy,
Rare-earth}
\maketitle

\section{Introduction}

The interface between flying and stationary qubits is emerging as a
key for progress in quantum information science. Impressive
steps have been made toward quantum computing using both atomic
\cite{pellizzari95,cirac95,monroe95} and photonic qubits \cite{turchette95,mattle95}. The ability to reversibly
transfer quantum states from single photons to single atoms would
enable progress in both of these areas. From the point of view of optical
quantum computing it would enable the single photon sources and
memories they require \cite{kuhn02}. From the point of view of quantum computing
using atoms it provides a potentially scalable way of effecting qubit-qubit
interactions \cite{cirac97,kimble08}.

The strong coupling regime of cavity quantum electrodynamics (cavity QED),
that is the regime where an atom-cavity interaction in some way
dominates the dissipation processes, is a necessity in order to enable the reversible transfer of
states between atoms and photons. This regime is achieved by having an atom coupled to an electromagnetic
field mode in a high-Q cavity. At present, strong coupling has been achieved experimentally using trapped
 atoms interacting with whispering gallery mode (WGM) \cite{aoki06,dayan08}, and Fabry-Perot
cavities \cite{boca04,boozer06,boozer07,colombe07, mckeever03,trupke07,wilk07}. It has also been
observed in the solid state using quantum dots in WGM \cite{srinivasan07}, micropillar \cite{reithmaier04} and
photonic crystal cavities \cite{hennessy07, yoshie04}.

Rare earth ion dopants are interesting systems to look at with respect
to cavity QED. Like other solid state systems embedded in a dielectric
they are particularly amenable to whispering gallery mode and
microstructured resonators. The highest intensity for the mode functions of the resonances of
such resonators are, in simple situations, inside the dielectric rather than
outside. Compared to the difficulties of trapping atoms or ions in
evanescent fields, the absolute position stability offered by
state optical centers is also attractive.

Rare earth ion dopants are at the same time much `cleaner' than other
solid state systems that have been investigated. They have very long
coherence times for their hyperfine transitions \cite{fraval04}, among
the longest of any qubit systems investigated. This would enable long
term storage of photon states \cite{longdell05}. They do not suffer from
the same spectral diffusion problems as quantum dots \cite{neuhauser00, sousa03} and do not have
the strong vibronic coupling of the nitrogen vacancy (NV) center \cite{davies76}.

Another reason for investigating cavity QED with rare earths is that
even if it is not possible to extend far into the strong coupling regime, the
cavity is able to help with the detection of single dopant ions.
 Rare earth ion
dopants are particularity attractive systems for quantum
computing. The long coherence times contrast strongly with large
ion-ion interactions. These interactions are due
to the fact that ions have static
electric dipole moments and these dipole moments are different in the
ground state and the excited state. The interaction strength for nearest
neighbors is in the region of GHz \cite{ohlsson02}. Furthermore the large ratio of
inhomogeneous to homogeneous broadening allows closely spaced ions to
be easily addressed by tuning the exciting laser \cite{ichimura01}. These properties
have all been measured and rudimentary demonstrations using ensembles
made \cite{pryde00,ichimura01,ohlsson02,longdell04}. While work has been done into improving the
scalability using ensembles \cite{roos04,wesenberg07}, and alternative
architectures such as ``read out ions'' \cite{nilsson04,rippe05,wesenberg07}, the
spin selective detection of a single rare earth ion dopant would represent a significant
step forward for rare earth ion based quantum computing.

There are two important issues that need to be considered with regards to
rare earth ions and cavity QED. Firstly the optical transitions have such long
coherence times because their oscillator strengths are weak. This means that the atom-cavity
coupling will be small, making it very difficult to achieve the `good cavity' strong coupling regime.
Secondly the atomic decay rate can be several orders of magnitude larger than the 
spontaneous emission rate calculated only from the oscillator
strength of the transition to a certain level. This is due to spontaneous emission to a level other than
the ground state followed by rapid non-radiative decay.

There have been a number of investigations into rare earth ion dopants and optical cavities.
Wang et al. \cite{wang98} investigated a monolithic Fabry-Perot resonator as a way of
improving the efficiency of photon echoes used for classical signal processing. Ichimura
and Goto \cite{ichimura06} observed optical bistability and normal-mode splitting in a
Pr:\YSO monolithic Fabry-Perot resonator, and concluded that they were close to achieving
the single atom strong coupling regime. Grudinin et al. \cite{grudinin06} did calculations
for Sm$^{2+}$ ions in CaF$_2$ microcavities, and concluded that based on the properties of
the free Sm$^{2+}$ ion, the strong coupling regime should be possible. We are concerned with
trivalent rare earth ions which have superior coherence properties.

Here we discuss briefly the strong coupling regime for ideal weak
oscillators and how this picture should change for the case of
realistic decay processes. We survey a number of rare earth ion doped
systems and discuss their utility in strong coupling cavity QED
experiments. We conclude by investigating some quantum information
operations in the bad cavity regime. We show that the transfer of
quantum states between atoms in distant cavities \cite{cirac97} is
still possible and introduce a new method for calculating the driving
fields needed for ``quantum impedance matching''. We also show that
phase shifts on single photons conditional on the state of single
atoms \cite{duan04} is possible, and suggest a heterodyne based
single atom detection scheme based on this.

\section{the meaning of strong coupling}

\begin{figure}[t]
\begin{center}
\includegraphics[scale=1]{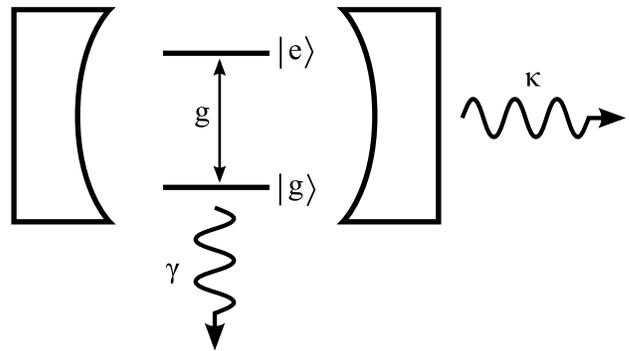}
\end{center}
\caption{Two-level atom in a cavity showing the three predominant interactions; g -
the coupling between the atom and cavity, $\kappa$ - the rate of decay out of the
cavity, $\gamma$ - the rate of decay through spontaneous emission of the atom. \label{fig:twolevelcavity}}
\end{figure}

We first consider a single two-level atom interacting with a single
cavity mode and a continuum of other optical modes that will be
treated as a bath. We will assume that the atom is sitting at an antinode
of the cavity mode field.  We will first examine the
ideal case where the only relaxation process of the atom is due to
its interaction with the optical fields. The dynamics of the system
can be described by three rates: the coupling between a single
atom and a single photon ($g$),
the cavity decay rate ($\kappa$) and the atom spontaneous emission rate
($\gamma$). Figure \ref{fig:twolevelcavity} is a diagram of this model showing the
interactions that occur.

From these constants one can derive two dimensionless numbers; the
critical atom number ($N_0$), which describes the
number of atoms required to have an appreciable effect on the cavity field, and the saturation photon number ($n_0$),
the number of photons required to saturate an atom in the cavity.
$N_0$ and $n_0$ are related to the system parameters by the following expressions:

\begin{equation}\label{eq:n_0}
N_0 \equiv \frac{\gamma\kappa}{g^2} \quad \mbox{and} \quad n_0 \equiv \frac{\gamma^2}{8g^2}
\end{equation}

We define the strong coupling regime as being when both the critical
atom number and the saturation photon number are less than one,
i.e. $(N_0, n_0)<1$. By looking at equation (\ref{eq:n_0}) we can see
that there are two ways that this condition can be achieved. The first
is the `good cavity' regime, when the atom-field coupling is
greater than the cavity and atomic decay rates
i.e. $g>(\kappa,\gamma)$. Weak oscillator strengths make $g$ small
making it difficult to achieve $g>\kappa$. Fortunately weak oscillators
also lead to small $\gamma$, and while $g$ scales linearly with the
transition dipole moment, the spontaneous emission rate scales as the
transition dipole moment squared. So for weak oscillators one
generally operates in the `bad cavity' regime, where the cavity decay rate is larger than the coupling
strength, but hopefully the atomic decay rate is still small enough to make
$N_0<1$.

For a single atom interacting with a single cavity mode we have the
following  interaction quantum Langevin equations \cite{gard85,Book-quantumnoise}:

\begin{eqnarray}
\dot{a} &=& g\sigma_- - \kappa a - \sqrt{2\kappa} b_{in}(t) \\
\dot{a}^{\dagger} &=& g\sigma_+ - \kappa a^{\dagger} - \sqrt{2\kappa} b_{in}^{\dagger}(t) \\ \label{eq:sigma-time}
\dot{\sigma}_- &=& 2 g\sigma_z a - \gamma \sigma_z \sigma_- + 2 \sqrt{\gamma} \sigma_z d_{in}(t) \\ \label{eq:sigma+time}
\dot{\sigma}_+ &=& 2 ga^{\dagger}\sigma_z - \gamma \sigma_+ \sigma_z + 2\sqrt{\gamma} \sigma_z d_{in}^{\dagger}(t) \\
\label{eq:sigmaztime}
\dot{\sigma}_z &=& -g(a^{\dagger} \sigma_- + \sigma_+ a) -\gamma \sigma_z -\frac{\gamma}{2} \\  \nonumber
&&+\sqrt{\gamma}(\sigma_+ d_{in}(t) + \sigma_- d_{in}^{\dagger}(t))
\end{eqnarray}

where $b_{in}(t)$ and $d_{in}(t)$ are the input fields associated with the cavity decay and
spontaneous emission respectively.  Assuming that $\kappa$ is much larger than the other
timescales, we adiabatically eliminate the cavity mode by setting
$\dot{a}=\dot{a}^{\dagger}=0$:

\begin{equation}\label{eq:anotime}
a=\frac{g\sigma_- -\sqrt{2\kappa} b_{in}(t)}{\kappa}
\end{equation}

Substituting (\ref{eq:anotime}) into (\ref{eq:sigma-time}) and
(\ref{eq:sigmaztime}) gives for $\sigma_-$ and $\sigma_z$:

\begin{eqnarray}
\dot{\sigma}_- &=& 2\left(\frac{g^2}{\kappa} +\frac{\gamma}{2}\right) \sigma_z \sigma_-
 - 2 \sqrt{\frac{2 g^2}{\kappa}} \sigma_z b_{in}(t) \\ \nonumber
&&+ 2\sqrt{\gamma} \sigma_z d_{in}(t) \\
\dot{\sigma}_z &=& -2 \left(\frac{g^2}{\kappa} + \frac{\gamma}{2}\right) \sigma_z - \left(\frac{g^2}{\kappa} + \frac{\gamma}{2}\right) \\ \nonumber
&&+ \sqrt{\frac{2 g^2}{\kappa}} ( \sigma_+ b_{in} (t) + \sigma_- b_{in}^{\dagger}(t)) \\ \nonumber
&&+ \sqrt{\gamma} ( \sigma_+ d_{in} (t) + \sigma_- d_{in}^{\dagger}(t))
\end{eqnarray}

We can see that after adiabatically eliminating the cavity, the atoms are
now described by two decay rates; $\frac{\gamma}{2}$ and $\frac{g^2}{\kappa}$.
$\frac{\gamma}{2}$ represents the spontaneous emission of the atom into free
space, and $\frac{g^2}{\kappa}$ represents the spontaneous emission of the
atom through a cavity field mode. So in the regime where the
critical atom number is small, ($N_0\ll 1$) the atom will decay through
the cavity rather than into free space.

\section{Realistic Decay Processes}

For a perfect two-level atom the spontaneous emission rate can be calculated from the 
oscillator strength of the transition. In reality there are electronic levels present 
other than the two involved in the transition. The atom can spontaneously emit to one of these other levels, then relax
the small energy gap to the ground state through much faster phonon decay
processes \cite{Book-spectroscofsolidsrareearth}. This process can cause the atomic 
decay rate to be several orders of magnitude larger than the theoretical two-level 
spontaneous emission rate, as the decay rate now has to be calculated by summing 
the contributions from all of the allowed transitions. This decay from the excited state
to other energy levels, followed by phonon decay does not have any effect on the
dynamics of the atom except to cause the atomic decay rate ($\gamma$)
to be larger.
In particular the nuclear spin state of the atom is preserved because the
decay is much more rapid than timescales over which the hyperfine
interaction is appreciable.
As a result the
effect of this branched decay path does not change the ideal atom picture, except
that the spontaneous emission rate $\gamma$ may be disappointingly large if we expect
it to be based only on the oscillator strength for the transition coupled to the cavity.

As well as the modification of $\gamma$ because of decay to other
energy levels, there is also broadening of the homogeneous linewidth
($\gamma_{h}$) due to excess dephasing. This dephasing, due to time
dependent frequency shifts of the atoms, can be caused by processes
such as  excitation or relaxation of other ions, the
nuclear and electron spins of the host material, and phonon scattering
in the ion \cite{equall95}. This last mechanism is temperature dependent and becomes
negligible at cryogenic temperatures for the right material
systems. In the case of cavity QED experiments the dominant cause of
excess dephasing will most likely be interactions between spins in the
host and the electron or nuclear spin of the dopant, as both the concentration
of dopant and the amount of optical excitation in the sample will be small.
Finding a good host for rare earth ion dopants that does not
contain nuclear spins has proved difficult, even in the quietest of
hosts such as Y$_2$SiO$_5$, the nuclear spins of the yttrium
contribute significantly to the homogeneous linewidths
\cite{equall94,equall95,macfarlane97}.

It has proved possible to ``turn off'' the excess dephasing due to
nuclear spins in the hosts in hyperfine transitions in Pr:\YSO
\cite{fraval04,fraval05} by working at a specific field where the
transition has zero first order Zeeman dependence. As well as
hyperfine transitions, this technique should also be applicable to
optical transitions in systems without electron spin, as long as the
nuclear spin of the dopant is large enough ($>1/2$) to provide a required anti-crossing.

Here we will  use $\gamma$ for the population decay rate ($\gamma=1/T_1$). The
homogeneous linewidth $\gamma_{h}=1/T_2$ is the sum of linewidth due to
population decay and excess dephasing ($\gamma_{p}$):

\begin{equation}
 \gamma_{h} = \frac{\gamma}{2} + \gamma_{p}
\end{equation}

The treatment of the previous section can now be applied to the more
realistic system that includes excess dephasing. We model the excess
dephasing with an interaction Hamiltonian that couples the atom's
population to the momentum of a bath of ground state harmonic
oscillators:

\begin{equation}
  H_\text{int}=i\hbar \sqrt{\frac{\gamma_p}{\pi}} \int  \sigma_z(f^\dagger(\omega)-f(\omega))d\omega
\end{equation}

where the $f(\omega)$ are the bath operators. We then follow the approach of Gardiner and Collett \cite{gard85,Book-quantumnoise}.

We will again assume that
 we are in the regime where the
cavity decays on a much faster timescale than dephasing of the
dopant ($\kappa > \gamma_h$). Again by adiabatically eliminating the cavity mode
($\dot{a}=\dot{a}^{\dagger} = 0$) we derive the quantum Langevin
equations:

\begin{eqnarray}
\dot{\sigma}_- &=& 2 \left(\frac{g^2}{\kappa} + \frac{\gamma}{2} + \gamma_p \right) \sigma_z \sigma_-  \\ \nonumber
 &&- 2 \sqrt{\frac{2 g^2}{\kappa}} \sigma_z b_{in}(t) + 2\sqrt{\gamma} \sigma_z d_{in}(t) \\ \nonumber
&&+ \sqrt{2 \gamma_p} \sigma_- (f_{in}^{\dagger}(t) -f_{in}(t)) \\
\dot{\sigma}_z &=& -2 \left(\frac{g^2}{\kappa} + \frac{\gamma}{2}\right) \sigma_z - \left(\frac{g^2}{\kappa} + \frac{\gamma}{2}\right) \\ \nonumber
&&+ \sqrt{\frac{2 g^2}{\kappa}} ( \sigma_+ b_{in} (t) + \sigma_- b_{in}^{\dagger}(t)) \\ \nonumber
&&+ \sqrt{\gamma} ( \sigma_+ d_{in} (t) + \sigma_- d_{in}^{\dagger}(t))
\end{eqnarray}

where $f_{in}(t)$ is the input field associated with the bath operator $f(\omega)$. 
We can see that dephasing adds an extra term to the equation
for $\dot{\sigma}_-$ which causes $\sigma_-$ to now be described by
the rates $\frac{g^2}{\kappa}$ and $\gamma_h$ (instead of
$\frac{\gamma}{2}$) which affects the coherence of the system. On the
other hand $\dot{\sigma_z}$ is unchanged, which means that the population
will still decay at the same rates as without excess dephasing.

So when dealing with cavity QED systems in the `bad cavity' limit and with
excess dephasing, the system should be parametrized by two different critical
atom numbers:

\begin{equation}
 N_{0(pop)} = \frac{\gamma \kappa}{g^2} \quad , \quad  N_{0(ph)} = \frac{2 \gamma_{h} \kappa}{g^2}
\end{equation}

Where because $\gamma_h\ge \gamma/2$ the phase critical atom number is
necessarily larger than the population critical atom number. A small
value for the population critical atom number will ensure that if
the dopant is excited the spontaneously emitted photon will
predominantly be emitted from the cavity. If a single photon pulse in a
single spatio-temporal mode or the coherent properties of the Fourier
transform limited single photon pulse are required, or if one desires
to probe the coherent properties of the atom cavity system, then
a small phase critical atom number is also necessary.

\section{Oscillators strengths Cavity parameters}

We are interested in using WGM cavities for achieving strong coupling with
rare earth ions because they have the ability to have small mode volumes ($V$)
while at the same time having high quality factors ($Q$)
\cite{vahala03, gorodetsky00}. We shall show that this ability to simultaneously
have small V and high Q is required in order to achieve strong coupling in such
a system.

Crystalline WGM resonators created using mechanical grinding and polishing techniques
have achieved some of the highest quality factors of any small optical cavity, with Q
factors up to $5.3\times10^{10}$ being reported in CaF$_2$ resonators \cite{grudinin06}.
As coherent spectroscopy of rare earth ion dopants is typically performed using doped crystals
\cite{Book-spectroscofsolidsrareearth}, crystalline WGM resonators are the
logical choice for achieving strong coupling in such systems.
 Rare earth doped glasses are
another option for cavity QED experiments, as it is relatively simple to manufacture glass microsphere
cavities with high quality factors (up to $8\times10^{9}$ in fused silica \cite{gorodetsky96}).
The problem with using doped glasses is that the homogeneous linewidth of the ions is several
orders of magnitude larger than for crystalline hosts \cite{Book-spectroscofsolidsrareearth},
which would make it very difficult to achieve strong coupling.

In order to reach the strong coupling regime we require the
atom-cavity coupling to be large while the atomic and cavity decay
rates are small. We here investigate how these rates specifically
relate to rare earth doped cavities, in particular WGM cavities.  The
cavity decay rate is given by \cite{buck03}:

\begin{equation} \label{eq:kappa}
\kappa = \frac{\pi c}{\lambda Q}
\end{equation}

where $\lambda$ is the wavelength of light in the cavity.
The coupling strength is given by the equation \cite{ichimura06}:

\begin{equation} \label{eq:g_0}
g = \frac{\mu}{n} \sqrt\frac{\omega_a}{2 \hbar \epsilon_0 V}
\end{equation}

where $n$ is the refractive index of the cavity and $\omega_a$ is the
frequency of the transition. From this equation we can see that in order
to have large coupling we require the cavity mode volume to be small.
The transition dipole moment ($\mu$) of the atom can be calculated from
the atom's oscillator strength (f) using the expression \cite{Book-rareearth}:

\begin{equation}\label{eq:dipole}
\mu^2 =\frac{3 \hbar e^2 n f}{2 m_e \omega \chi_L}
\end{equation}

where $\chi_L=((n^2+2)/3)^2$ is the local correction to the electric field to account
for the fact that the ion is less polarisable than the bulk medium \cite{Book-rareearth}.

The transitions between 4f levels in the rare earths are of particular
interest when looking at cavity QED applications due to their long population
and coherence lifetimes. Having long lifetimes means that the atomic decay
rates are very small, which makes it easier to get into a strong coupling regime
(because $\gamma \propto f$ and $g \propto \sqrt{f}$, as explained in section II).
Having small atomic  decay rates makes it easy to get into a regime
where ($\gamma,\gamma_{h}$) $<g$. This will enable the use of larger
cavities than are traditionally used for cavity QED experiments, as
$g \propto \frac{1}{\sqrt{V}}$. Using millimetre sized resonators
is desirable for two main reasons: 1) they are easy to fabricate (can be made by
hand using a standard lathe), 2) they can be tuned easily by deformation of
the resonator.
From equation (\ref{eq:n_0}) we can
see that having a very small $\gamma$ means that it is easy to achieve
$n_0<1$. So when dealing with rare earths the main issue is having
$\kappa$ small enough (high Q factor) that $N_0<1$.

The spontaneous emission time $T_{spon}$ is the time it would
take for the excited state of the atom to relax to the ground state if it
was just a simple two-level atom with no other energy levels present.
$T_{spon}$ is related to the transition dipole moment of the atom \cite{Book-rareearth}:

\begin{equation}\label{eq:tspon}
T_{spon} = \frac{3 \epsilon_0 \hbar \lambda^3}{8 \pi^2 n \chi_L \mu^2}
\end{equation}

In terms of the properties of the rare earth ion transition ($T_1$, $T_2$, $T_{spon}$) we can write
the critical atom and saturation photon numbers as:

\begin{eqnarray}
N_{0(pop)}& =& \frac{\gamma \kappa}{g^2} \\ \nonumber
  &=& \frac{\beta}{Q} \frac{T_{spon}}{T_1} \chi_L
\end{eqnarray}

\begin{eqnarray}
N_{0(ph)}& =& \frac{2 \gamma_{h} \kappa}{g^2} \\ \nonumber
  &=&  \frac{2 \beta}{Q} \frac{T_{spon}}{T_2} \chi_L
\end{eqnarray}

\begin{eqnarray}
n_0&=& \frac{\gamma \gamma_{h}}{4 g^2} \\ \nonumber
  &=& \frac{\lambda \beta}{4 \pi c} \frac{T_{spon}}{T_1 T_2} \chi_L
\end{eqnarray}

where we have introduced a new dimensionless parameter $\beta$ that
describes the mode volume of the cavity compared to the wavelength
cubed \cite{buck03}.:

\begin{equation}
 \beta = \frac{8 \pi^2 n^3 V}{3 \lambda^{3}}
\end{equation}

Table~\ref{fig:QEDparams} shows parameters for a number of rare earth
ion systems. Purely by looking at the ratios of $\frac{T_{spon}}{T_2}$ and
$\frac{T_{spon}}{T_1}$ we would say that the $^3$H$_{6}$ - $^3$H$_{4}$
transition in Tm$^{3+}$:LiNbO$_3$ would be the best option for
investigating strong coupling, while the $^4$I$_{9/2}$ -
$^4$F$_{3/2}$ transition in Nd$^{3+}$:YVO$_4$ and the $^4$I$_{15/2}$ -
$^4$I$_{13/2}$ transition in Er$^{3+}$:Y$_2$SiO$_5$ are also
possibilities. The $^3$H$_4$ - $^1$D$_2$ transition in Pr$^{3+}$:YAG is
another likely candidate as it has a small $\frac{T_{spon}}{T_1}$
ratio.  The $\frac{T_{spon}}{T_2}$ ratio is not as good as the others,
but by applying magnetic field to the system it should be possible to
lengthen the dephasing time.

\begin{table*}[t]
\footnotesize

\begin{center}\label{fig:QEDparams}
\begin{tabular}{|c|c|c|c|c|c|c|c|c|}
\hline
Transition & $\lambda$(nm) & Oscillator Strength & $\mu$($10^{-32}$ (Cm)) & $T_{spon}$(ms) & $T_1$($\mu$s) & $T_2$($\mu$s) & $\frac{T_{spon}}{T_1}$ &  $\frac{T_{spon}}{T_2}$ \\
\hline

$^3$H$_4$ - $^1$D$_2$ in Pr$^{3+}$:Y$_2$SiO$_5$ &605.977 \cite{wang97} & $3 \times 10^{-7}$ \cite{equall95} & 1.59 &5.66 &164 \cite{equall95} &152 (77G)\cite{equall95} &34.5 &37.2 \\

$^3$H$_4$ - $^1$D$_2$ in Pr$^{3+}$:YAG &609.587\cite{Book-rareearth} &$1.5 \times 10^{-6}$ \cite{pryagnote} &3.53 &1.11 &230\cite{Book-rareearth} &20 (zero field)\cite{Book-rareearth} &4.83 &55.5 \\

$^4$I$_{9/2}$ - $^4$F$_{3/2}$ in Nd$^{3+}$:YVO$_4$ &879.705\cite{Book-rareearth} &$8 \times 10^{-6}$\cite{Book-rareearth} &9.16 &0.366 &100\cite{riedmatten08} &27 (15 kG) \cite{equall95} &3.66 &13.6 \\

$^4$I$_{15/2}$ - $^4$I$_{13/2}$ in Er$^{3+}$:Y$_2$SiO$_5$ &1536.14\cite{Book-rareearth} & $2\times 10^{-7}$ \cite{thielsuncone} & 2.07& 54.6&11400\cite{Book-rareearth} &4080 (70kG)\cite{Book-rareearth} & 4.79 & 13.4 \\

$^4$I$_{15/2}$ - $^4$I$_{13/2}$ in Er$^{3+}$:LiNbO$_3$ &1531.52\cite{Book-rareearth} & $8\times 10^{-7}$ \cite{thielsuncone}& 3.50 & 9.08 &2000\cite{thielsuncone} &80 (20kG) \cite{thielsuncone} &4.54 &113.5 \\

$^3$H$_{6}$ - $^3$H$_{4}$ in Tm$^{3+}$:LiNbO$_3$ &794.264 \cite{thielsuncone} & $5.044 \times 10^{-6}$ \cite{wang96} &6.37 &0.382 &170\cite{thielsuncone} &32 (200G)\cite{thielsuncone} &2.25 &11.9 \\

$^3$H$_{6}$ - $^3$H$_{4}$ in Tm$^{3+}$:YAG &793.156 \cite{Book-rareearth} &$6.3 \times 10^{-8}$ \cite{Book-rareearth} &0.824&44.6 &800\cite{Book-rareearth} &130 (100G)\cite{Book-rareearth} &55.8 &343\\

$^7$F$_{0}$ - $^5$D$_{0}$ in Eu$^{3+}$:Y$_2$SiO$_5$ &579.879\cite{Book-rareearth} & $1.3 \times 10^{-8}$\cite{konz03} &0.324 &120 &1900\cite{Book-rareearth} &2600 (100G)\cite{Book-rareearth} &63.2 &46.2\\

\hline
\end{tabular}
\end{center}
\caption{Parameters of Rare Earth Transitions. The magnetic field
  values next to each $T_2$ corresponds to the magnetic field used for
  the measurement.}
\end{table*}

\begin{figure}[t]
\begin{center}
\includegraphics[width=0.48\textwidth]{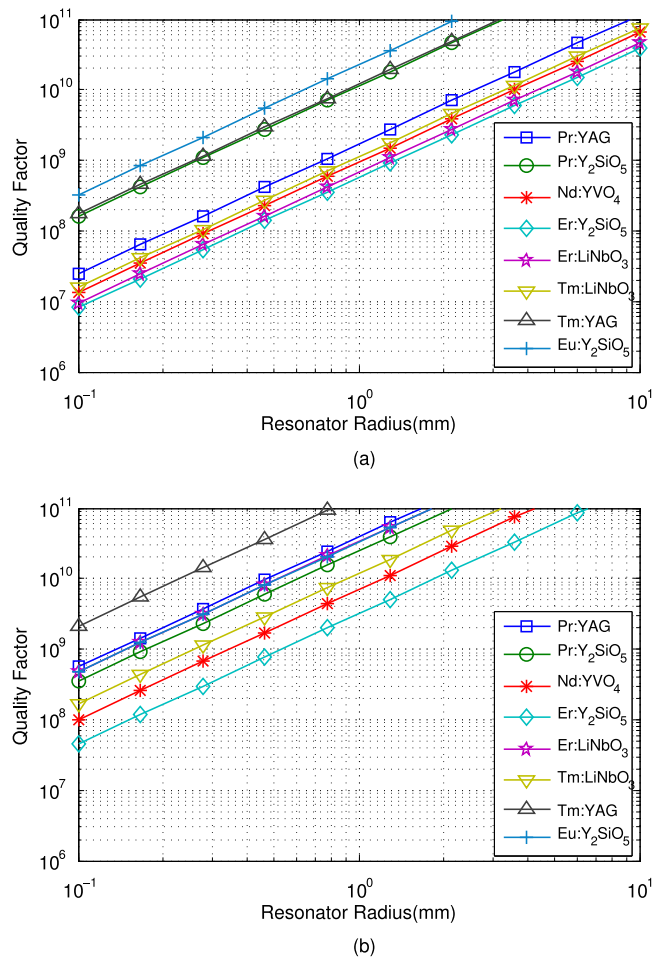}
\end{center}
\caption{The resonator radius and quality factor required for a)
$N_{0(pop)}=1$, b) $N_{0(ph)}=1$ for different rare earth ions coupled
to the the fundamental mode in a WGM resonator. \label{fig:rvQ}}
\end{figure}

To see how the ion parameters relate to WGM cavities, in
Fig. \ref{fig:rvQ}  we have plotted the cavity radius versus quality
factor when $(N_{0(pop)}, N_{0(ph)})=1$ for the different
transitions using the fundamental TM mode ($n=1, m=\ell$) of a
spherical cavity.
We can see that Er$^{3+}$:Y$_2$SiO$_5$ is
clearly the best, although it does have the disadvantage of needing large
magnetic fields to achieve good $T_2$ values. Nd$^{3+}$:YVO$_4$ and
Tm$^{3+}$:LiNbO$_3$ also appear to be good candidates. These
plots show that wavelength plays quite a large part in determining how
suitable a transition is for strong coupling, which we would expect due
to the $\frac{1}{\lambda^3}$ dependence of $\beta$.

\section{Throw and catch}

One of the great promises of cavity QED is the ability to use
light to transfer quantum states between the metastable ground
states of two different atoms \cite{cirac97}. Here we show this
to be possible in the bad cavity regime and introduce a new
method for calculating which pulse shape is required from the
coupling beams in order to achieve the ``quantum impedance matching''.

\begin{figure}[t]
\begin{center}
\includegraphics[scale=1]{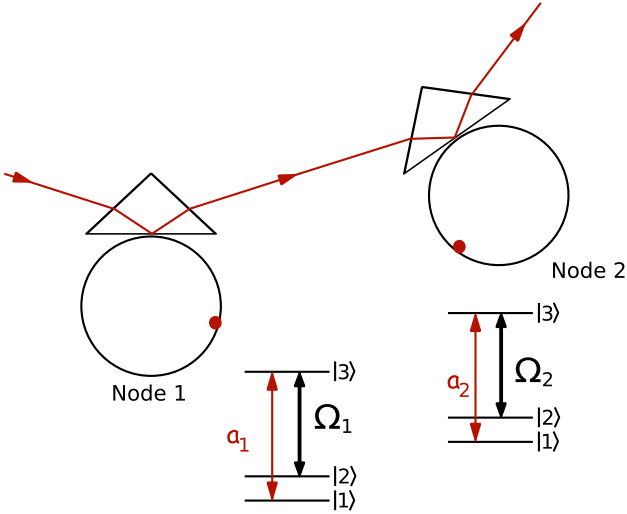}
\end{center}
\caption{Throw and catch of photons between two atoms in separate
  cavities. The $a_1$ and $a_2$ represent cavity modes and $\Omega_1$
  and $\Omega_2$ classical driving fields, perhaps introduced from
the side of the cavity. \label{fig:throw_catch}}
\end{figure}

Here we consider two lambda systems in two separate
cavities with the output of one cavity driving the other as shown in Fig. \ref{fig:throw_catch}.

We start with the Hamiltonian for a single lambda system in a cavity:

\begin{equation}
H=g(a^{\dagger} \sigma_{13} + a \sigma_{31}) + \Omega(t) (\sigma_{32} +\sigma_{23})
\end{equation}

Following a similar procedure to section II we derive a set of
Quantum Langevin equations that describe the time evolution:

\begin{eqnarray}
\dot{\sigma}_{13} &=& -i  g a(\sigma_{11} -\sigma_{33}) - i \Omega(t) \sigma_{12} \\
\dot{\sigma}_{12} &=& i g a \sigma_{32} - i \Omega(t) \sigma_{13} \\
\dot{a} &=& -i g \sigma_{13} - \kappa a - \sqrt{2} \kappa b_{in}(t)
\end{eqnarray}

where $\sigma_{13} = |1\rangle \langle3|$, etc.

Because we are interested in the case of at most one excitation of
this system, it is sufficient for us to consider the amplitudes given by:

\begin{eqnarray}
\phi_{12} &=& \langle vac,1|\sigma_{12}|\psi \rangle \\
\phi_{13} &=& \langle vac,1|\sigma_{13}|\psi \rangle \\
\alpha &=& \langle vac,1|a|\psi \rangle
\end{eqnarray}

where $\langle vac,1|$ refers to the cavity being in the vacuum state and the atom being in state 1.
This leads to equations of motion:
\begin{equation}
\dot{\rho}_1=\mathcal{L}_1 \rho_1
\end{equation}

where:

\begin{equation}
\mathbf{\rho_1} =
\left( \begin{array}{c}
\alpha \\
\phi_{12} \\
\phi_{13}
\end{array} \right)\quad,\quad \mathbf{\mathcal{L}_1} =
\left( \begin{array}{ccc}
-\kappa & 0 & -i g \\
0 & 0 & -i \Omega_1(t) \\
-i g & - i \Omega_1(t) & 0
\end{array} \right)
\end{equation}

Now suppose we arrange it so that the output of system 1 drives the
input of system 2 (which is identical to system 1), i.e. $b_{in} =
a_{out}$ then we can write a combined set of equations of motion for
the entire system:

\begin{equation}
\frac{d}{dt}
\left( \begin{array}{c}
\rho_1 \\
\rho_2
\end{array} \right) = \left( \begin{array}{cc}
\mathcal{L}_1 & 0 \\
X & \mathcal{L}_2
\end{array} \right)\left( \begin{array}{c}
\rho_1 \\
\rho_2
\end{array} \right) \quad \mbox{where} \quad X=
\left( \begin{array}{ccc}
2 \kappa & 0 & 0 \\
0 & 0 & 0 \\
0& 0 & 0
\end{array} \right)
\end{equation}

Solving these equations will give an expression for $\Omega(t)$
showing how the Rabi frequency of the atoms needs to be driven in
order to achieve a pulse being emitted at the first node which
propagates and is then received at the second node (hence the throw
and catch). To do this we write the equation of motion for the state
vector $\rho$ as:

\begin{equation}
\dot{\rho} = A\rho + \Omega(t) B \rho + \sqrt{2 \kappa} \left( \begin{array}{c}
\beta_{in}(t) \\
0 \\
0
\end{array} \right)
\end{equation}

where:

\begin{equation}
A =
\left( \begin{array}{ccc}
-\kappa & 0 & -i g \\
0 & 0 & 0\\
-i g & 0 & 0
\end{array} \right)\quad,\quad B =
\left( \begin{array}{ccc}
0 & 0 & 0\\
0 & 0 & -i  \\
0 & - i  & 0
\end{array} \right)
\end{equation}

We can take the input-output relation for the system:

\begin{equation}
\beta_{out} = \beta_{in} + \sqrt{2 \kappa} [ 1 0 0 ] \rho
\end{equation}

Differentiate twice with respect to $t$:

\begin{eqnarray}
\dot{\beta}_{out} &=& \dot{\beta}_{in} + 2 \kappa \beta_{in} + \sqrt{2 \kappa} [1 0 0 ] A \rho \\
\ddot{\beta}_{out} &=& \ddot{\beta}_{in} + 2 \kappa \dot{\beta}_{in} + \sqrt{2 \kappa} [1 0 0 ] A^2 \rho \nonumber \\
&&+ \sqrt{2 \kappa} [1 0 0 ] A B \rho \Omega(t) + 2 \kappa A(1,1) \beta_{in}
\end{eqnarray}

This can then be rearranged to get an expression for the required Rabi
driving field in terms of the input and output fields:

\begin{equation}
\Omega(t) = \frac{ \ddot{\beta}_{out} - \ddot{\beta}_{in} - 2 \kappa \dot{\beta}_{in} - 2 \kappa A(1,1) \beta_{in} - \sqrt{2 \kappa} [1 0 0 ] A^2 \rho}    {\sqrt{2 \kappa} [1 0 0 ] A B \rho}
\end{equation}

\begin{figure}[t]
\begin{center}
\includegraphics[width=0.48\textwidth]{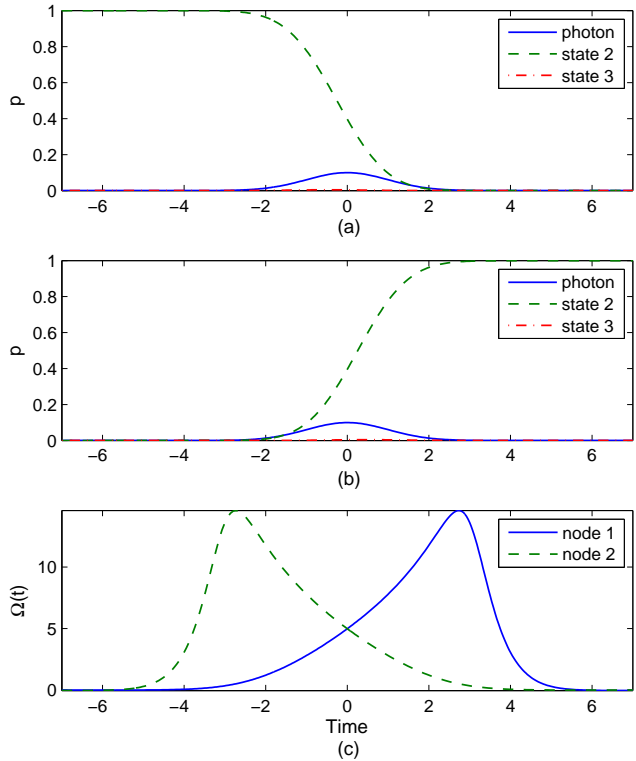}
\end{center}
\caption{The results of simulation of quantum state transfer in the
  bad cavity regime. Graphs (a) and (b) show the atom and cavity
  populations for nodes 1 and 2 respectively. Graph (c) shows the
  classical driving fields applied to each node. 
 The atom at node 1 starts off in state $\ket{2}$ and as the driving field is applied
it gets transferred via a Ramen process to state $\ket{1}$, emitting a photon into the cavity mode in the process. 
This photon is then completely absorbed at node 2
placing that atom in state $\ket{2}$. Note that the units of time and $\Omega(t)$ are
arbitrary and have been set to fit the parameters used in the simulation ($g=10$, $\kappa = 2$ and $\gamma=0$).
\label{fig:throwandcatch}}
\end{figure}

Now we have an expression for the Rabi driving field that can be used
to model the transfer of information between two nodes. This is done
by setting the output field of the first node as a Gaussian and
calculating the $\Omega(t)$ required to achieve this output.
We then use the fact that absorbtion of a photon is the time
reverse of emitting a photon to get the driving field at node 2, which
is just the time reversed field at node 1.
The results of a simulation showing successful quantum state transfer
is shown in Fig.~\ref{fig:throwandcatch}.

\section{Conditional phase shifts and single dopant detection}

One way to use the optical cavity to enable single dopant detection is
as a collecting lens for the fluorescence and then photon
counting. This has a number of benefits over a conventional collecting
lens: firstly the lens is narrow band meaning that it will only collect fluorescence
over its linewidth, and secondly in the strong coupling regime the
effective decay rate of the atom ($g^2/\kappa$) is increased leading
to higher count rates. As in general the cavity will be resonant with
the transition from one particular ground state hyperfine level to one
particular excited state hyperfine level, working in the strong
coupling regime will improve the cyclicity of the transition. This is
important as except for complicated schemes based on parity
conservation \cite{longdell03} it is difficult to find a cyclic
transition in rare earth ion dopants.

The emission rates will still be small, of the order of tens of
kilohertz. This along with the poor cyclic nature of the rare earth
ion transitions, and finite dark count rates of single photon
detectors, will make single state readout difficult.

Here we propose a single atom detection scheme based on coherent
detection, inspired by the work of Wrigge et al. \cite{wrigge07}
and a quantum controlled phase-flip gate as described by Duan and
Kimble \cite{duan04}.

Duan and Kimble proposed a quantum gate between a photon and the hyperfine
state of an atom in a one-sided cavity. If the atom was in a state not
involved in the transition coupled to the cavity mode, then the cavity
was in effect empty and a single photon would be reflected from the
cavity. If the atom was in a state involved in the transition coupled
to the cavity mode, and if the system was in the good cavity strong
coupling regime, then the effect of the atom was to split the cavity
resonance in two and an incident photon would be reflected from the
cavity with a $\pi$ phase shift, much as it would be if the rear
mirror of the cavity were not present.

Here we show that this conditional phase shift is still present in the
bad cavity limit, even though the vacuum Rabi splitting is small
compared to the cavity linewidth. This enables the presence of the
atom to be detected as a narrow band phase shift on a weak coherent
incident field.

To keep the phase shift as large as possible, we will assume that the
input field is kept weak enough so as not to saturate the atom; in
this situation the atom can be treated as a harmonic oscillator
leading to quantum Langevin equations:

\begin{equation}
 \begin{bmatrix}
  \dot{a}\\\dot{s}\\
 \end{bmatrix}
=
 \begin{bmatrix}
  -\kappa -i\delta &+g\\-g&-\gamma/2-i\delta \\
 \end{bmatrix}
\begin{bmatrix}
  a\\s\\
 \end{bmatrix}
-
 \begin{bmatrix}
  \sqrt{2\kappa}&0 \\
  0&\sqrt{\gamma}\\
 \end{bmatrix}
\begin{bmatrix}
  a_{in}(t)\\s_{in}(t)\\
 \end{bmatrix}
\end{equation}

\begin{figure}[t]
  \centering
  \includegraphics[width=0.48\textwidth]{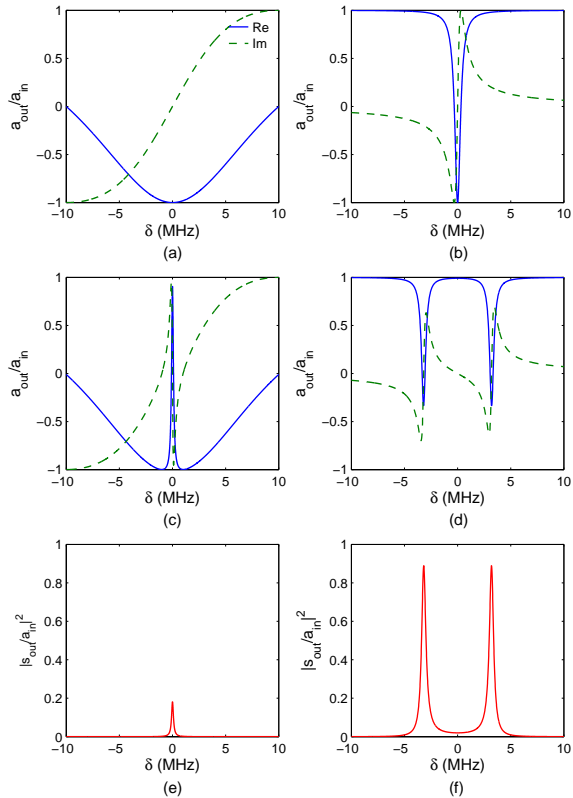}
\caption{\label{fig:phase_shift} Phase of the cavity output as a
function of detuning for (a),(b) an empty cavity; (c),(d) a cavity
with an atom in it. Note that the cavity output at zero detuning
undergoes a $\pi$  phase shift when there is a resonant atom
present. (e)\&(f) show the probability of the atom spontaneously
emitting a photon vs detuning. The left plots correspond to an
atom-cavity system in the bad cavity regime with parameters $(g,\kappa,\gamma)=(1,10,0.01)$MHz.
The right plots are for a system in the good cavity regime with parameters $(g,\kappa,\gamma)=(3.2,0.32,0.32)$MHz.}
\end{figure}

\begin{figure}[t]
  \centering
  \includegraphics[width=0.48\textwidth]{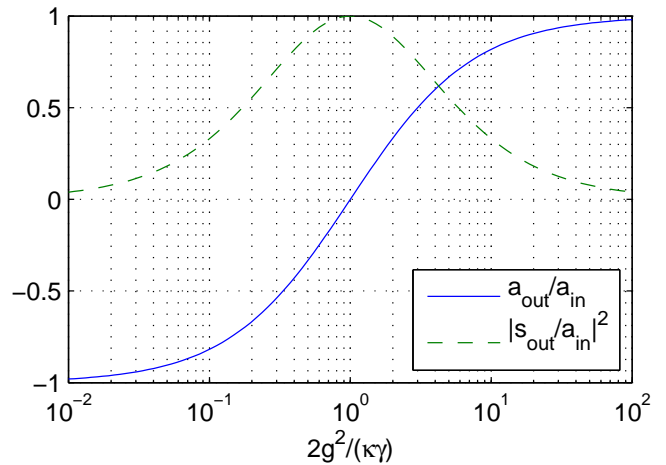}
\caption{\label{fig:phase_shift2} Phase shift and spontaneous emission
probability vs $2g^2/\kappa \gamma$ for the case of zero detuning. For
small values of $g$ the atom is essentially uncoupled from the cavity,
the input light is phase shifted by $\pi$. For $2g^2/\kappa
\gamma=1$ there is no light output from the cavity and all the input
light is lost as spontaneous emission. For large vales of $g$ the
incident light is phase shifted and the atom not excited, enabling the
presence of the atom to be determined non-destructively.}
\end{figure}

Here $\delta$ is the detuning between the input field and the cavity,
$a$ and $s$ are the lowering operators for the cavity and atom
respectively. We shall assume the cavity is resonant with the
dopant. We can calculate the spectral response of such a system if we
assume the input field is narrow band compared to the
dynamics of the atom cavity system. Setting $\dot{a}(t)=\dot{s}(t)=0$  leads to:
\begin{equation}
  \begin{bmatrix}
  a\\s\\
 \end{bmatrix}
=
\text{Inv}\left(
 \begin{bmatrix}
  -\kappa -i\delta &+g\\-g&-\gamma/2-i\delta \\
 \end{bmatrix}
\right)
\begin{bmatrix}
  \sqrt{2\kappa}&0 \\
  0&\sqrt{\gamma}\\
 \end{bmatrix}
\begin{bmatrix}
  a_{in}(t)\\s_{in}(t)\\
\end{bmatrix}
\end{equation}
which along with the input output relations gives:

\begin{widetext}
\begin{eqnarray}
\begin{bmatrix}
  a_{out}(t)\\s_{out}(t)\\
\end{bmatrix}
= \frac{1}{D}
\begin{bmatrix}
 g^2+(i\delta+\gamma/2)(i\delta-\kappa)&-\sqrt{2}g\sqrt{\kappa\gamma}\\
\sqrt{2}g\sqrt{\kappa\gamma}&g^2+(i\delta-\gamma/2)(i\delta+\kappa)\\
\end{bmatrix}
\begin{bmatrix}
  a_{in}(t)\\s_{in}(t)\\
\end{bmatrix}
\end{eqnarray}
\end{widetext}

where $D = g^2+(i\delta+\gamma/2)(i\delta+\kappa)$. Plots of this
behaviour is shown in Figs. \ref{fig:phase_shift} and \ref{fig:phase_shift2}.

Having an atom resonant in the cavity (Fig. \ref{fig:phase_shift} (c),(d))
causes a phase shift of $\pi$ compared to the empty cavity case
(Fig. \ref{fig:phase_shift} (a),(b)). This shift occurs
 independent of whether we are in the good or bad cavity regime.
Fig. \ref{fig:phase_shift} (e),(f) show that in both regimes there is still
a probability of the atom spontaneously emitting a photon, but the further
we get into the strong coupling regime (i.e. the smaller $N_0$ is), the smaller
this probability becomes, as shown in Fig. \ref{fig:phase_shift2}.

Coherent detection of the atom in this manner moves the problem from
single photon counting with low count rates to the detection of
narrow bandwidth phase shifts. In order to not be limited by the
exciting laser its linewidth and drift should be small compared to the
ions. State of the art stable lasers have both short and long term stability
much greater than the kilohertz linewidths of rare earth ions \cite{alnis08}.
One benefit of the coherent approach is that the
quantum efficiency of detectors used for homodyne and heterodyne
detection is higher than photon counting detectors. The coherent
approach will also be very insensitive to stray light, due to the spatial
selectivity of the homodyne/heterodyne detection coupled with very
small detection bandwidth.  For these reasons the coherent approach
should have both  increased sensitivity and selectivity over photon counting.

In practice these narrowband phase shifts could be detected like an
optical free induction decay. A brief, weak pulse exciting the
atom-cavity system would lead to long-lived coherent emission at the
same frequency as the ion. If this brief pulse was created as a phase
modulated sideband from an electro-optic modulator with the carrier
light off-resonance from the atom-cavity system, no interferometer
would be needed, simplifying the implementation. In this way the
approach could be considered as cavity-enhanced FM spectroscopy
\cite{bjorklund80}.

\section{Conclusion}
We have investigated the use of rare earth ions in cavity QED and determined that by using
crystalline WGM cavities it should be possible to achieve the `bad cavity' strong coupling
regime. We have showed that by operating in the bad cavity regime, quantum states can be
reversibly transferred between atoms in separate cavities, which is an important requirement
if we wish to use rare earth doped cavities for quantum information processing. And finally
we have devised a method of detecting single atoms by using an optical cavity and measuring
the phase shift of photons that interact with this atom-cavity system.


\begin{thebibliography}{60}
\expandafter\ifx\csname natexlab\endcsname\relax\def\natexlab#1{#1}\fi
\expandafter\ifx\csname bibnamefont\endcsname\relax
  \def\bibnamefont#1{#1}\fi
\expandafter\ifx\csname bibfnamefont\endcsname\relax
  \def\bibfnamefont#1{#1}\fi
\expandafter\ifx\csname citenamefont\endcsname\relax
  \def\citenamefont#1{#1}\fi
\expandafter\ifx\csname url\endcsname\relax
  \def\url#1{\texttt{#1}}\fi
\expandafter\ifx\csname urlprefix\endcsname\relax\def\urlprefix{URL }\fi
\providecommand{\bibinfo}[2]{#2}
\providecommand{\eprint}[2][]{\url{#2}}

\bibitem[{\citenamefont{{Pellizzari} et~al.}(1995)\citenamefont{{Pellizzari},
  {Gardiner}, {Cirac}, and {Zoller}}}]{pellizzari95}
\bibinfo{author}{\bibfnamefont{T.}~\bibnamefont{{Pellizzari}}},
  \bibinfo{author}{\bibfnamefont{S.~A.} \bibnamefont{{Gardiner}}},
  \bibinfo{author}{\bibfnamefont{J.~I.} \bibnamefont{{Cirac}}},
  \bibnamefont{and} \bibinfo{author}{\bibfnamefont{P.}~\bibnamefont{{Zoller}}},
  \bibinfo{journal}{Physical Review Letters} \textbf{\bibinfo{volume}{75}},
  \bibinfo{pages}{3788} (\bibinfo{year}{1995}).

\bibitem[{\citenamefont{Cirac and Zoller}(1995)}]{cirac95}
\bibinfo{author}{\bibfnamefont{J.~I.} \bibnamefont{Cirac}} \bibnamefont{and}
  \bibinfo{author}{\bibfnamefont{P.}~\bibnamefont{Zoller}},
  \bibinfo{journal}{Phys. Rev. Lett.} \textbf{\bibinfo{volume}{74}},
  \bibinfo{pages}{4091} (\bibinfo{year}{1995}).

\bibitem[{\citenamefont{Monroe et~al.}(1995)\citenamefont{Monroe, Meekhof,
  King, Itano, and Wineland}}]{monroe95}
\bibinfo{author}{\bibfnamefont{C.}~\bibnamefont{Monroe}},
  \bibinfo{author}{\bibfnamefont{D.~M.} \bibnamefont{Meekhof}},
  \bibinfo{author}{\bibfnamefont{B.~E.} \bibnamefont{King}},
  \bibinfo{author}{\bibfnamefont{W.~M.} \bibnamefont{Itano}}, \bibnamefont{and}
  \bibinfo{author}{\bibfnamefont{D.~J.} \bibnamefont{Wineland}},
  \bibinfo{journal}{Phys. Rev. Lett.} \textbf{\bibinfo{volume}{75}},
  \bibinfo{pages}{4714} (\bibinfo{year}{1995}).

\bibitem[{\citenamefont{Turchette et~al.}(1995)\citenamefont{Turchette, Hood,
  Lange, Mabuchi, and Kimble}}]{turchette95}
\bibinfo{author}{\bibfnamefont{Q.~A.} \bibnamefont{Turchette}},
  \bibinfo{author}{\bibfnamefont{C.~J.} \bibnamefont{Hood}},
  \bibinfo{author}{\bibfnamefont{W.}~\bibnamefont{Lange}},
  \bibinfo{author}{\bibfnamefont{H.}~\bibnamefont{Mabuchi}}, \bibnamefont{and}
  \bibinfo{author}{\bibfnamefont{H.~J.} \bibnamefont{Kimble}},
  \bibinfo{journal}{Phys. Rev. Lett.} \textbf{\bibinfo{volume}{75}},
  \bibinfo{pages}{4710} (\bibinfo{year}{1995}).

\bibitem[{\citenamefont{Mattle et~al.}(1996)\citenamefont{Mattle, Weinfurter,
  Kwiat, and Zeilinger}}]{mattle95}
\bibinfo{author}{\bibfnamefont{K.}~\bibnamefont{Mattle}},
  \bibinfo{author}{\bibfnamefont{H.}~\bibnamefont{Weinfurter}},
  \bibinfo{author}{\bibfnamefont{P.~G.} \bibnamefont{Kwiat}}, \bibnamefont{and}
  \bibinfo{author}{\bibfnamefont{A.}~\bibnamefont{Zeilinger}},
  \bibinfo{journal}{Phys. Rev. Lett.} \textbf{\bibinfo{volume}{76}},
  \bibinfo{pages}{4656} (\bibinfo{year}{1996}).

\bibitem[{\citenamefont{Kuhn et~al.}(2002)\citenamefont{Kuhn, Hennrich, and
  Rempe}}]{kuhn02}
\bibinfo{author}{\bibfnamefont{A.}~\bibnamefont{Kuhn}},
  \bibinfo{author}{\bibfnamefont{M.}~\bibnamefont{Hennrich}}, \bibnamefont{and}
  \bibinfo{author}{\bibfnamefont{G.}~\bibnamefont{Rempe}},
  \bibinfo{journal}{Phys. Rev. Lett.} \textbf{\bibinfo{volume}{89}},
  \bibinfo{pages}{067901} (\bibinfo{year}{2002}).

\bibitem[{\citenamefont{Cirac et~al.}(1997)\citenamefont{Cirac, Zoller, Kimble,
  and Mabuchi}}]{cirac97}
\bibinfo{author}{\bibfnamefont{J.~I.} \bibnamefont{Cirac}},
  \bibinfo{author}{\bibfnamefont{P.}~\bibnamefont{Zoller}},
  \bibinfo{author}{\bibfnamefont{H.~J.} \bibnamefont{Kimble}},
  \bibnamefont{and} \bibinfo{author}{\bibfnamefont{H.}~\bibnamefont{Mabuchi}},
  \bibinfo{journal}{Phys. Rev. Lett.} \textbf{\bibinfo{volume}{78}},
  \bibinfo{pages}{3221} (\bibinfo{year}{1997}).

\bibitem[{\citenamefont{{Kimble}}(2008)}]{kimble08}
\bibinfo{author}{\bibfnamefont{H.~J.} \bibnamefont{{Kimble}}},
  \bibinfo{journal}{\nat} \textbf{\bibinfo{volume}{453}}, \bibinfo{pages}{1023}
  (\bibinfo{year}{2008}).

\bibitem[{\citenamefont{Aoki et~al.}(2006)\citenamefont{Aoki, Dayan, Wilcut,
  Bowen., Parkins, Kippenberg, Vahala, and Kimble}}]{aoki06}
\bibinfo{author}{\bibfnamefont{T.}~\bibnamefont{Aoki}},
  \bibinfo{author}{\bibfnamefont{B.}~\bibnamefont{Dayan}},
  \bibinfo{author}{\bibfnamefont{E.}~\bibnamefont{Wilcut}},
  \bibinfo{author}{\bibfnamefont{W.~P.} \bibnamefont{Bowen.}},
  \bibinfo{author}{\bibfnamefont{A.~S.} \bibnamefont{Parkins}},
  \bibinfo{author}{\bibfnamefont{T.~J.} \bibnamefont{Kippenberg}},
  \bibinfo{author}{\bibfnamefont{K.~J.} \bibnamefont{Vahala}},
  \bibnamefont{and} \bibinfo{author}{\bibfnamefont{H.~J.}
  \bibnamefont{Kimble}}, \bibinfo{journal}{Nature}
  \textbf{\bibinfo{volume}{443}}, \bibinfo{pages}{671} (\bibinfo{year}{2006}).

\bibitem[{\citenamefont{{Dayan} et~al.}(2008)\citenamefont{{Dayan}, {Parkins},
  {Aoki}, {Ostby}, {Vahala}, and {Kimble}}}]{dayan08}
\bibinfo{author}{\bibfnamefont{B.}~\bibnamefont{{Dayan}}},
  \bibinfo{author}{\bibfnamefont{A.~S.} \bibnamefont{{Parkins}}},
  \bibinfo{author}{\bibfnamefont{T.}~\bibnamefont{{Aoki}}},
  \bibinfo{author}{\bibfnamefont{E.~P.} \bibnamefont{{Ostby}}},
  \bibinfo{author}{\bibfnamefont{K.~J.} \bibnamefont{{Vahala}}},
  \bibnamefont{and} \bibinfo{author}{\bibfnamefont{H.~J.}
  \bibnamefont{{Kimble}}}, \bibinfo{journal}{Science}
  \textbf{\bibinfo{volume}{319}}, \bibinfo{pages}{1062} (\bibinfo{year}{2008}).

\bibitem[{\citenamefont{Boca et~al.}(2004)\citenamefont{Boca, Miller, Birnbaum,
  Boozer, McKeever, and Kimble}}]{boca04}
\bibinfo{author}{\bibfnamefont{A.}~\bibnamefont{Boca}},
  \bibinfo{author}{\bibfnamefont{R.}~\bibnamefont{Miller}},
  \bibinfo{author}{\bibfnamefont{K.~M.} \bibnamefont{Birnbaum}},
  \bibinfo{author}{\bibfnamefont{A.~D.} \bibnamefont{Boozer}},
  \bibinfo{author}{\bibfnamefont{J.}~\bibnamefont{McKeever}}, \bibnamefont{and}
  \bibinfo{author}{\bibfnamefont{H.~J.} \bibnamefont{Kimble}},
  \bibinfo{journal}{Physical Review Letters} \textbf{\bibinfo{volume}{93}},
  \bibinfo{pages}{233603} (\bibinfo{year}{2004}).

\bibitem[{\citenamefont{{Boozer} et~al.}(2006)\citenamefont{{Boozer}, {Boca},
  {Miller}, {Northup}, and {Kimble}}}]{boozer06}
\bibinfo{author}{\bibfnamefont{A.~D.} \bibnamefont{{Boozer}}},
  \bibinfo{author}{\bibfnamefont{A.}~\bibnamefont{{Boca}}},
  \bibinfo{author}{\bibfnamefont{R.}~\bibnamefont{{Miller}}},
  \bibinfo{author}{\bibfnamefont{T.~E.} \bibnamefont{{Northup}}},
  \bibnamefont{and} \bibinfo{author}{\bibfnamefont{H.~J.}
  \bibnamefont{{Kimble}}}, \bibinfo{journal}{Physical Review Letters}
  \textbf{\bibinfo{volume}{97}}, \bibinfo{pages}{083602}
  (\bibinfo{year}{2006}).

\bibitem[{\citenamefont{{Boozer} et~al.}(2007)\citenamefont{{Boozer}, {Boca},
  {Miller}, {Northup}, and {Kimble}}}]{boozer07}
\bibinfo{author}{\bibfnamefont{A.~D.} \bibnamefont{{Boozer}}},
  \bibinfo{author}{\bibfnamefont{A.}~\bibnamefont{{Boca}}},
  \bibinfo{author}{\bibfnamefont{R.}~\bibnamefont{{Miller}}},
  \bibinfo{author}{\bibfnamefont{T.~E.} \bibnamefont{{Northup}}},
  \bibnamefont{and} \bibinfo{author}{\bibfnamefont{H.~J.}
  \bibnamefont{{Kimble}}}, \bibinfo{journal}{Physical Review Letters}
  \textbf{\bibinfo{volume}{98}}, \bibinfo{pages}{193601}
  (\bibinfo{year}{2007}).

\bibitem[{\citenamefont{{Colombe} et~al.}(2007)\citenamefont{{Colombe},
  {Steinmetz}, {Dubois}, {Linke}, {Hunger}, and {Reichel}}}]{colombe07}
\bibinfo{author}{\bibfnamefont{Y.}~\bibnamefont{{Colombe}}},
  \bibinfo{author}{\bibfnamefont{T.}~\bibnamefont{{Steinmetz}}},
  \bibinfo{author}{\bibfnamefont{G.}~\bibnamefont{{Dubois}}},
  \bibinfo{author}{\bibfnamefont{F.}~\bibnamefont{{Linke}}},
  \bibinfo{author}{\bibfnamefont{D.}~\bibnamefont{{Hunger}}}, \bibnamefont{and}
  \bibinfo{author}{\bibfnamefont{J.}~\bibnamefont{{Reichel}}},
  \bibinfo{journal}{\nat} \textbf{\bibinfo{volume}{450}}, \bibinfo{pages}{272}
  (\bibinfo{year}{2007}).

\bibitem[{\citenamefont{{McKeever} et~al.}(2003)\citenamefont{{McKeever},
  {Boca}, {Boozer}, {Buck}, and {Kimble}}}]{mckeever03}
\bibinfo{author}{\bibfnamefont{J.}~\bibnamefont{{McKeever}}},
  \bibinfo{author}{\bibfnamefont{A.}~\bibnamefont{{Boca}}},
  \bibinfo{author}{\bibfnamefont{A.~D.} \bibnamefont{{Boozer}}},
  \bibinfo{author}{\bibfnamefont{J.~R.} \bibnamefont{{Buck}}},
  \bibnamefont{and} \bibinfo{author}{\bibfnamefont{H.~J.}
  \bibnamefont{{Kimble}}}, \bibinfo{journal}{\nat}
  \textbf{\bibinfo{volume}{425}}, \bibinfo{pages}{268} (\bibinfo{year}{2003}).

\bibitem[{\citenamefont{{Trupke} et~al.}(2007)\citenamefont{{Trupke},
  {Goldwin}, {Darqui{\'e}}, {Dutier}, {Eriksson}, {Ashmore}, and
  {Hinds}}}]{trupke07}
\bibinfo{author}{\bibfnamefont{M.}~\bibnamefont{{Trupke}}},
  \bibinfo{author}{\bibfnamefont{J.}~\bibnamefont{{Goldwin}}},
  \bibinfo{author}{\bibfnamefont{B.}~\bibnamefont{{Darqui{\'e}}}},
  \bibinfo{author}{\bibfnamefont{G.}~\bibnamefont{{Dutier}}},
  \bibinfo{author}{\bibfnamefont{S.}~\bibnamefont{{Eriksson}}},
  \bibinfo{author}{\bibfnamefont{J.}~\bibnamefont{{Ashmore}}},
  \bibnamefont{and} \bibinfo{author}{\bibfnamefont{E.~A.}
  \bibnamefont{{Hinds}}}, \bibinfo{journal}{Physical Review Letters}
  \textbf{\bibinfo{volume}{99}}, \bibinfo{pages}{063601}
  (\bibinfo{year}{2007}).

\bibitem[{\citenamefont{{Wilk} et~al.}(2007)\citenamefont{{Wilk}, {Webster},
  {Kuhn}, and {Rempe}}}]{wilk07}
\bibinfo{author}{\bibfnamefont{T.}~\bibnamefont{{Wilk}}},
  \bibinfo{author}{\bibfnamefont{S.~C.} \bibnamefont{{Webster}}},
  \bibinfo{author}{\bibfnamefont{A.}~\bibnamefont{{Kuhn}}}, \bibnamefont{and}
  \bibinfo{author}{\bibfnamefont{G.}~\bibnamefont{{Rempe}}},
  \bibinfo{journal}{Science} \textbf{\bibinfo{volume}{317}},
  \bibinfo{pages}{488} (\bibinfo{year}{2007}).

\bibitem[{\citenamefont{Srinivasan and Painter}(2007)}]{srinivasan07}
\bibinfo{author}{\bibfnamefont{K.}~\bibnamefont{Srinivasan}} \bibnamefont{and}
  \bibinfo{author}{\bibfnamefont{O.}~\bibnamefont{Painter}},
  \bibinfo{journal}{\nat} \textbf{\bibinfo{volume}{450}}, \bibinfo{pages}{862}
  (\bibinfo{year}{2007}).

\bibitem[{\citenamefont{{Reithmaier} et~al.}(2004)\citenamefont{{Reithmaier},
  {S\c{e}k}, {L{\"o}ffler}, {Hofmann}, {Kuhn}, {Reitzenstein}, {Keldysh},
  {Kulakovskii}, {Reinecke}, and {Forchel}}}]{reithmaier04}
\bibinfo{author}{\bibfnamefont{J.~P.} \bibnamefont{{Reithmaier}}},
  \bibinfo{author}{\bibfnamefont{G.}~\bibnamefont{{S\c{e}k}}},
  \bibinfo{author}{\bibfnamefont{A.}~\bibnamefont{{L{\"o}ffler}}},
  \bibinfo{author}{\bibfnamefont{C.}~\bibnamefont{{Hofmann}}},
  \bibinfo{author}{\bibfnamefont{S.}~\bibnamefont{{Kuhn}}},
  \bibinfo{author}{\bibfnamefont{S.}~\bibnamefont{{Reitzenstein}}},
  \bibinfo{author}{\bibfnamefont{L.~V.} \bibnamefont{{Keldysh}}},
  \bibinfo{author}{\bibfnamefont{V.~D.} \bibnamefont{{Kulakovskii}}},
  \bibinfo{author}{\bibfnamefont{T.~L.} \bibnamefont{{Reinecke}}},
  \bibnamefont{and}
  \bibinfo{author}{\bibfnamefont{A.}~\bibnamefont{{Forchel}}},
  \bibinfo{journal}{\nat} \textbf{\bibinfo{volume}{432}}, \bibinfo{pages}{197}
  (\bibinfo{year}{2004}).

\bibitem[{\citenamefont{{Hennessy} et~al.}(2007)\citenamefont{{Hennessy},
  {Badolato}, {Winger}, {Gerace}, {Atat{\"u}re}, {Gulde}, {F{\"a}lt}, {Hu}, and
  {Imamo\u{g}lu}}}]{hennessy07}
\bibinfo{author}{\bibfnamefont{K.}~\bibnamefont{{Hennessy}}},
  \bibinfo{author}{\bibfnamefont{A.}~\bibnamefont{{Badolato}}},
  \bibinfo{author}{\bibfnamefont{M.}~\bibnamefont{{Winger}}},
  \bibinfo{author}{\bibfnamefont{D.}~\bibnamefont{{Gerace}}},
  \bibinfo{author}{\bibfnamefont{M.}~\bibnamefont{{Atat{\"u}re}}},
  \bibinfo{author}{\bibfnamefont{S.}~\bibnamefont{{Gulde}}},
  \bibinfo{author}{\bibfnamefont{S.}~\bibnamefont{{F{\"a}lt}}},
  \bibinfo{author}{\bibfnamefont{E.~L.} \bibnamefont{{Hu}}}, \bibnamefont{and}
  \bibinfo{author}{\bibfnamefont{A.}~\bibnamefont{{Imamo\u{g}lu}}},
  \bibinfo{journal}{\nat} \textbf{\bibinfo{volume}{445}}, \bibinfo{pages}{896}
  (\bibinfo{year}{2007}).

\bibitem[{\citenamefont{{Yoshie} et~al.}(2004)\citenamefont{{Yoshie},
  {Scherer}, {Hendrickson}, {Khitrova}, {Gibbs}, {Rupper}, {Ell}, {Shchekin},
  and {Deppe}}}]{yoshie04}
\bibinfo{author}{\bibfnamefont{T.}~\bibnamefont{{Yoshie}}},
  \bibinfo{author}{\bibfnamefont{A.}~\bibnamefont{{Scherer}}},
  \bibinfo{author}{\bibfnamefont{J.}~\bibnamefont{{Hendrickson}}},
  \bibinfo{author}{\bibfnamefont{G.}~\bibnamefont{{Khitrova}}},
  \bibinfo{author}{\bibfnamefont{H.~M.} \bibnamefont{{Gibbs}}},
  \bibinfo{author}{\bibfnamefont{G.}~\bibnamefont{{Rupper}}},
  \bibinfo{author}{\bibfnamefont{C.}~\bibnamefont{{Ell}}},
  \bibinfo{author}{\bibfnamefont{O.~B.} \bibnamefont{{Shchekin}}},
  \bibnamefont{and} \bibinfo{author}{\bibfnamefont{D.~G.}
  \bibnamefont{{Deppe}}}, \bibinfo{journal}{\nat}
  \textbf{\bibinfo{volume}{432}}, \bibinfo{pages}{200} (\bibinfo{year}{2004}).

\bibitem[{\citenamefont{Fraval et~al.}(2004)\citenamefont{Fraval, Sellars, and
  Longdell}}]{fraval04}
\bibinfo{author}{\bibfnamefont{E.}~\bibnamefont{Fraval}},
  \bibinfo{author}{\bibfnamefont{M.~J.} \bibnamefont{Sellars}},
  \bibnamefont{and} \bibinfo{author}{\bibfnamefont{J.~J.}
  \bibnamefont{Longdell}}, \bibinfo{journal}{Phys. Rev. Lett.}
  \textbf{\bibinfo{volume}{92}}, \bibinfo{pages}{077601}
  (\bibinfo{year}{2004}).

\bibitem[{\citenamefont{Longdell et~al.}(2005)\citenamefont{Longdell, Fraval,
  Sellars, and Manson}}]{longdell05}
\bibinfo{author}{\bibfnamefont{J.~J.} \bibnamefont{Longdell}},
  \bibinfo{author}{\bibfnamefont{E.}~\bibnamefont{Fraval}},
  \bibinfo{author}{\bibfnamefont{M.~J.} \bibnamefont{Sellars}},
  \bibnamefont{and} \bibinfo{author}{\bibfnamefont{N.~B.}
  \bibnamefont{Manson}}, \bibinfo{journal}{Phys. Rev. Lett.}
  \textbf{\bibinfo{volume}{95}}, \bibinfo{pages}{063601}
  (\bibinfo{year}{2005}).

\bibitem[{\citenamefont{Neuhauser et~al.}(2000)\citenamefont{Neuhauser,
  Shimizu, Woo, Empedocles, and Bawendi}}]{neuhauser00}
\bibinfo{author}{\bibfnamefont{R.~G.} \bibnamefont{Neuhauser}},
  \bibinfo{author}{\bibfnamefont{K.~T.} \bibnamefont{Shimizu}},
  \bibinfo{author}{\bibfnamefont{W.~K.} \bibnamefont{Woo}},
  \bibinfo{author}{\bibfnamefont{S.~A.} \bibnamefont{Empedocles}},
  \bibnamefont{and} \bibinfo{author}{\bibfnamefont{M.~G.}
  \bibnamefont{Bawendi}}, \bibinfo{journal}{Physical Review Letters}
  \textbf{\bibinfo{volume}{85}}, \bibinfo{pages}{3301} (\bibinfo{year}{2000}).

\bibitem[{\citenamefont{{de Sousa} and {Das Sarma}}(2003)}]{sousa03}
\bibinfo{author}{\bibfnamefont{R.}~\bibnamefont{{de Sousa}}} \bibnamefont{and}
  \bibinfo{author}{\bibfnamefont{S.}~\bibnamefont{{Das Sarma}}},
  \bibinfo{journal}{Phys. Rev. B} \textbf{\bibinfo{volume}{68}},
  \bibinfo{pages}{115322} (\bibinfo{year}{2003}).

\bibitem[{\citenamefont{Davies and Hamer}(1976)}]{davies76}
\bibinfo{author}{\bibfnamefont{G.}~\bibnamefont{Davies}} \bibnamefont{and}
  \bibinfo{author}{\bibfnamefont{M.~F.} \bibnamefont{Hamer}},
  \bibinfo{journal}{Royal Society of London Proceedings Series A}
  \textbf{\bibinfo{volume}{348}}, \bibinfo{pages}{285} (\bibinfo{year}{1976}).

\bibitem[{\citenamefont{{N. Ohlsson} et~al.}(2002)\citenamefont{{N. Ohlsson},
  {Krishna Mohan}, and Kr{\"o}ll}}]{ohlsson02}
\bibinfo{author}{\bibfnamefont{N.}~\bibnamefont{{N. Ohlsson}}},
  \bibinfo{author}{\bibfnamefont{R.}~\bibnamefont{{Krishna Mohan}}},
  \bibnamefont{and}
  \bibinfo{author}{\bibfnamefont{S.}~\bibnamefont{Kr{\"o}ll}},
  \bibinfo{journal}{Optics Communications} \textbf{\bibinfo{volume}{201}},
  \bibinfo{pages}{71} (\bibinfo{year}{2002}).

\bibitem[{\citenamefont{Ichimura}(2001)}]{ichimura01}
\bibinfo{author}{\bibfnamefont{K.}~\bibnamefont{Ichimura}},
  \bibinfo{journal}{Optics Communications} \textbf{\bibinfo{volume}{196}},
  \bibinfo{pages}{119} (\bibinfo{year}{2001}).

\bibitem[{\citenamefont{Pryde et~al.}(2000)\citenamefont{Pryde, Sellars, and
  Manson}}]{pryde00}
\bibinfo{author}{\bibfnamefont{G.~J.} \bibnamefont{Pryde}},
  \bibinfo{author}{\bibfnamefont{M.~J.} \bibnamefont{Sellars}},
  \bibnamefont{and} \bibinfo{author}{\bibfnamefont{N.~B.}
  \bibnamefont{Manson}}, \bibinfo{journal}{Phys. Rev. Lett.}
  \textbf{\bibinfo{volume}{84}}, \bibinfo{pages}{1152} (\bibinfo{year}{2000}).

\bibitem[{\citenamefont{Longdell et~al.}(2004)\citenamefont{Longdell, Sellars,
  and Manson}}]{longdell04}
\bibinfo{author}{\bibfnamefont{J.~J.} \bibnamefont{Longdell}},
  \bibinfo{author}{\bibfnamefont{M.~J.} \bibnamefont{Sellars}},
  \bibnamefont{and} \bibinfo{author}{\bibfnamefont{N.~B.}
  \bibnamefont{Manson}}, \bibinfo{journal}{Physical Review Letters}
  \textbf{\bibinfo{volume}{93}}, \bibinfo{pages}{130503}
  (\bibinfo{year}{2004}).

\bibitem[{\citenamefont{Roos and M{\o}lmer}(2004)}]{roos04}
\bibinfo{author}{\bibfnamefont{I.}~\bibnamefont{Roos}} \bibnamefont{and}
  \bibinfo{author}{\bibfnamefont{K.}~\bibnamefont{M{\o}lmer}},
  \bibinfo{journal}{Phys. Rev. A} \textbf{\bibinfo{volume}{69}},
  \bibinfo{pages}{022321} (\bibinfo{year}{2004}).

\bibitem[{\citenamefont{Wesenberg et~al.}(2007)\citenamefont{Wesenberg,
  M{\o}lmer, Rippe, and Kr{\"o}ll}}]{wesenberg07}
\bibinfo{author}{\bibfnamefont{J.~H.} \bibnamefont{Wesenberg}},
  \bibinfo{author}{\bibfnamefont{K.}~\bibnamefont{M{\o}lmer}},
  \bibinfo{author}{\bibfnamefont{L.}~\bibnamefont{Rippe}}, \bibnamefont{and}
  \bibinfo{author}{\bibfnamefont{S.}~\bibnamefont{Kr{\"o}ll}},
  \bibinfo{journal}{\pra} \textbf{\bibinfo{volume}{75}},
  \bibinfo{pages}{012304} (\bibinfo{year}{2007}).

\bibitem[{\citenamefont{Nilsson et~al.}(2004)\citenamefont{Nilsson, Rippe,
  Kr{\"o}ll, Klieber, and Suter}}]{nilsson04}
\bibinfo{author}{\bibfnamefont{M.}~\bibnamefont{Nilsson}},
  \bibinfo{author}{\bibfnamefont{L.}~\bibnamefont{Rippe}},
  \bibinfo{author}{\bibfnamefont{S.}~\bibnamefont{Kr{\"o}ll}},
  \bibinfo{author}{\bibfnamefont{R.}~\bibnamefont{Klieber}}, \bibnamefont{and}
  \bibinfo{author}{\bibfnamefont{D.}~\bibnamefont{Suter}},
  \bibinfo{journal}{Phys. Rev. B} \textbf{\bibinfo{volume}{70}},
  \bibinfo{pages}{214116} (\bibinfo{year}{2004}).

\bibitem[{\citenamefont{Rippe et~al.}(2005)\citenamefont{Rippe, Nilsson,
  Kr{\"o}ll, Klieber, and Suter}}]{rippe05}
\bibinfo{author}{\bibfnamefont{L.}~\bibnamefont{Rippe}},
  \bibinfo{author}{\bibfnamefont{M.}~\bibnamefont{Nilsson}},
  \bibinfo{author}{\bibfnamefont{S.}~\bibnamefont{Kr{\"o}ll}},
  \bibinfo{author}{\bibfnamefont{R.}~\bibnamefont{Klieber}}, \bibnamefont{and}
  \bibinfo{author}{\bibfnamefont{D.}~\bibnamefont{Suter}},
  \bibinfo{journal}{Phys. Rev. A} \textbf{\bibinfo{volume}{71}},
  \bibinfo{pages}{062328} (\bibinfo{year}{2005}).

\bibitem[{\citenamefont{Wang et~al.}(1998)\citenamefont{Wang, Greiner, and
  Mossberg}}]{wang98}
\bibinfo{author}{\bibfnamefont{T.}~\bibnamefont{Wang}},
  \bibinfo{author}{\bibfnamefont{C.}~\bibnamefont{Greiner}}, \bibnamefont{and}
  \bibinfo{author}{\bibfnamefont{T.~W.} \bibnamefont{Mossberg}},
  \bibinfo{journal}{Optics Letters} \textbf{\bibinfo{volume}{23}},
  \bibinfo{pages}{1736} (\bibinfo{year}{1998}).

\bibitem[{\citenamefont{Ichimura and Goto}(2006)}]{ichimura06}
\bibinfo{author}{\bibfnamefont{K.}~\bibnamefont{Ichimura}} \bibnamefont{and}
  \bibinfo{author}{\bibfnamefont{H.}~\bibnamefont{Goto}},
  \bibinfo{journal}{Physical Review A} \textbf{\bibinfo{volume}{74}},
  \bibinfo{pages}{033818} (\bibinfo{year}{2006}).

\bibitem[{\citenamefont{Grudinin et~al.}(2006)\citenamefont{Grudinin, Matsko,
  Savchenkov, Strekalov, Ilchenko, and Maleki}}]{grudinin06}
\bibinfo{author}{\bibfnamefont{I.~S.} \bibnamefont{Grudinin}},
  \bibinfo{author}{\bibfnamefont{A.~B.} \bibnamefont{Matsko}},
  \bibinfo{author}{\bibfnamefont{A.~A.} \bibnamefont{Savchenkov}},
  \bibinfo{author}{\bibfnamefont{D.}~\bibnamefont{Strekalov}},
  \bibinfo{author}{\bibfnamefont{V.~S.} \bibnamefont{Ilchenko}},
  \bibnamefont{and} \bibinfo{author}{\bibfnamefont{L.}~\bibnamefont{Maleki}},
  \bibinfo{journal}{Optics Communications} \textbf{\bibinfo{volume}{265}},
  \bibinfo{pages}{33} (\bibinfo{year}{2006}).

\bibitem[{\citenamefont{Duan and Kimble}(2004)}]{duan04}
\bibinfo{author}{\bibfnamefont{L.~M.} \bibnamefont{Duan}} \bibnamefont{and}
  \bibinfo{author}{\bibfnamefont{H.~J.} \bibnamefont{Kimble}},
  \bibinfo{journal}{Physical Review Letters} \textbf{\bibinfo{volume}{92}},
  \bibinfo{pages}{127902} (\bibinfo{year}{2004}).


\bibitem[{\citenamefont{Gardeiner and Collett}(1985)}]{gard85}
\bibinfo{author}{\bibfnamefont{C.~W.} \bibnamefont{Gardiner}} \bibnamefont{and}
  \bibinfo{author}{\bibfnamefont{M.~J.} \bibnamefont{Collett}},
  \bibinfo{journal}{Physical Review A} \textbf{\bibinfo{volume}{31}},
  \bibinfo{pages}{3761-3774} (\bibinfo{year}{1985}).

\bibitem[{\citenamefont{Gardiner and Zoller}(2000)}]{Book-quantumnoise}
\bibinfo{author}{\bibfnamefont{C.~W.} \bibnamefont{Gardiner}} \bibnamefont{and}
  \bibinfo{author}{\bibfnamefont{P.}~\bibnamefont{Zoller}},
  \emph{\bibinfo{title}{{Quantum Noise (2nd Edition)}}}
  (\bibinfo{publisher}{Springer-Verlag, Berlin Heidelberg},
  \bibinfo{year}{2000}).

\bibitem[{\citenamefont{{Equall} et~al.}(1995)\citenamefont{{Equall}, {Cone},
  and {Macfarlane}}}]{equall95}
\bibinfo{author}{\bibfnamefont{R.~W.} \bibnamefont{{Equall}}},
  \bibinfo{author}{\bibfnamefont{R.~L.} \bibnamefont{{Cone}}},
  \bibnamefont{and} \bibinfo{author}{\bibfnamefont{R.~M.}
  \bibnamefont{{Macfarlane}}}, \bibinfo{journal}{Physical Review B}
  \textbf{\bibinfo{volume}{52}}, \bibinfo{pages}{3963} (\bibinfo{year}{1995}).

\bibitem[{\citenamefont{Equall et~al.}(1994)\citenamefont{Equall, Sun, Cone,
  and Macfarlane}}]{equall94}
\bibinfo{author}{\bibfnamefont{R.~W.} \bibnamefont{Equall}},
  \bibinfo{author}{\bibfnamefont{Y.}~\bibnamefont{Sun}},
  \bibinfo{author}{\bibfnamefont{R.~L.} \bibnamefont{Cone}}, \bibnamefont{and}
  \bibinfo{author}{\bibfnamefont{R.~M.} \bibnamefont{Macfarlane}},
  \bibinfo{journal}{Phys. Rev. Lett.} \textbf{\bibinfo{volume}{72}},
  \bibinfo{pages}{2179} (\bibinfo{year}{1994}).

\bibitem[{\citenamefont{Macfarlane et~al.}(1997)\citenamefont{Macfarlane,
  Harris, Sun, Cone, and Equall}}]{macfarlane97}
\bibinfo{author}{\bibfnamefont{R.~M.} \bibnamefont{Macfarlane}},
  \bibinfo{author}{\bibfnamefont{T.~L.} \bibnamefont{Harris}},
  \bibinfo{author}{\bibfnamefont{Y.}~\bibnamefont{Sun}},
  \bibinfo{author}{\bibfnamefont{R.~L.} \bibnamefont{Cone}}, \bibnamefont{and}
  \bibinfo{author}{\bibfnamefont{R.~W.} \bibnamefont{Equall}},
  \bibinfo{journal}{Optics Letters} \textbf{\bibinfo{volume}{22}},
  \bibinfo{pages}{871} (\bibinfo{year}{1997}).

\bibitem[{\citenamefont{Fraval et~al.}(2005)\citenamefont{Fraval, Sellars, and
  Longdell}}]{fraval05}
\bibinfo{author}{\bibfnamefont{E.}~\bibnamefont{Fraval}},
  \bibinfo{author}{\bibfnamefont{M.~J.} \bibnamefont{Sellars}},
  \bibnamefont{and} \bibinfo{author}{\bibfnamefont{J.~J.}
  \bibnamefont{Longdell}}, \bibinfo{journal}{Phys. Rev. Lett.}
  \textbf{\bibinfo{volume}{95}}, \bibinfo{pages}{030506}
  (\bibinfo{year}{2005}).

\bibitem[{\citenamefont{Walls and Milburn}(1994)}]{Book-quantumoptics}
\bibinfo{author}{\bibfnamefont{D.~F.} \bibnamefont{Walls}} \bibnamefont{and}
  \bibinfo{author}{\bibfnamefont{G.~J.} \bibnamefont{Milburn}},
  \emph{\bibinfo{title}{{Quantum Optics}}}
  (\bibinfo{publisher}{Springer-Verlag, Berlin Heidelberg},
  \bibinfo{year}{1994}).

\bibitem[{\citenamefont{Vahala}(2003)}]{vahala03}
\bibinfo{author}{\bibfnamefont{K.~J.} \bibnamefont{Vahala}},
  \bibinfo{journal}{Nature} \textbf{\bibinfo{volume}{424}},
  \bibinfo{pages}{839} (\bibinfo{year}{2003}).

\bibitem[{\citenamefont{Gorodetsky et~al.}(2000)\citenamefont{Gorodetsky,
  Pryamikov, and Ilchenko}}]{gorodetsky00}
\bibinfo{author}{\bibfnamefont{M.~L.} \bibnamefont{Gorodetsky}},
  \bibinfo{author}{\bibfnamefont{A.~D.} \bibnamefont{Pryamikov}},
  \bibnamefont{and} \bibinfo{author}{\bibfnamefont{V.~S.}
  \bibnamefont{Ilchenko}}, \bibinfo{journal}{Journal of the Optical Society of
  America B} \textbf{\bibinfo{volume}{17}}, \bibinfo{pages}{1051}
  (\bibinfo{year}{2000}).

\bibitem[{\citenamefont{{Kaplyanskii} and
  {Macfarlane}}(1987)}]{Book-spectroscofsolidsrareearth}
\bibinfo{editor}{\bibfnamefont{A.~A.} \bibnamefont{{Kaplyanskii}}}
  \bibnamefont{and} \bibinfo{editor}{\bibfnamefont{R.~M.}
  \bibnamefont{{Macfarlane}}}, eds., \emph{\bibinfo{title}{{Spectroscopy of
  Solids Containing Rare Earth Ions}}} (\bibinfo{publisher}{North-Holland
  Physics Publishing, Amsterdam}, \bibinfo{year}{1987}).

\bibitem[{\citenamefont{Gorodetsky et~al.}(1996)\citenamefont{Gorodetsky,
  Savchenkov, and Ilchenko}}]{gorodetsky96}
\bibinfo{author}{\bibfnamefont{M.~L.} \bibnamefont{Gorodetsky}},
  \bibinfo{author}{\bibfnamefont{A.~A.} \bibnamefont{Savchenkov}},
  \bibnamefont{and} \bibinfo{author}{\bibfnamefont{V.~S.}
  \bibnamefont{Ilchenko}}, \bibinfo{journal}{Optics Letters}
  \textbf{\bibinfo{volume}{21}}, \bibinfo{pages}{453} (\bibinfo{year}{1996}).

\bibitem[{\citenamefont{Buck and Kimble}(2003)}]{buck03}
\bibinfo{author}{\bibfnamefont{J.~R.} \bibnamefont{Buck}} \bibnamefont{and}
  \bibinfo{author}{\bibfnamefont{H.~J.} \bibnamefont{Kimble}},
  \bibinfo{journal}{Physical Review A} \textbf{\bibinfo{volume}{67}},
  \bibinfo{pages}{033806} (\bibinfo{year}{2003}).

\bibitem[{\citenamefont{Liu and Jacquier}(2005)}]{Book-rareearth}
\bibinfo{editor}{\bibfnamefont{G.}~\bibnamefont{Liu}} \bibnamefont{and}
  \bibinfo{editor}{\bibfnamefont{B.}~\bibnamefont{Jacquier}}, eds.,
  \emph{\bibinfo{title}{{Spectroscopic Properties of Rare Earths in Optical
  Materials}}} (\bibinfo{publisher}{Tsinghua University Press and
  Springer-Verlag Berlin Heidelberg}, \bibinfo{year}{2005}).

\bibitem[{\citenamefont{Wang}(1997)}]{wang97}
\bibinfo{author}{\bibfnamefont{G.}~\bibnamefont{Wang}}, Ph.D. thesis,
  \bibinfo{school}{Montana State University} (\bibinfo{year}{1997}).

\bibitem[{pry()}]{pryagnote}
\bibinfo{note}{Calculated.}

\bibitem[{\citenamefont{{de Riedmatten} et~al.}(2008)\citenamefont{{de
  Riedmatten}, {Afzelius}, {Staudt}, {Simon}, and {Gisin}}}]{riedmatten08}
\bibinfo{author}{\bibfnamefont{H.}~\bibnamefont{{de Riedmatten}}},
  \bibinfo{author}{\bibfnamefont{M.}~\bibnamefont{{Afzelius}}},
  \bibinfo{author}{\bibfnamefont{M.}~\bibnamefont{{Staudt}}},
  \bibinfo{author}{\bibfnamefont{C.}~\bibnamefont{{Simon}}}, \bibnamefont{and}
  \bibinfo{author}{\bibfnamefont{N.}~\bibnamefont{{Gisin}}},
  \bibinfo{journal}{\nat} \textbf{\bibinfo{volume}{456}}, \bibinfo{pages}{773}
  (\bibinfo{year}{2008}).

\bibitem[{\citenamefont{Thiel et~al.}()\citenamefont{Thiel, Sun, and
  Cone}}]{thielsuncone}
\bibinfo{author}{\bibfnamefont{C.~W.} \bibnamefont{Thiel}},
  \bibinfo{author}{\bibfnamefont{Y.}~\bibnamefont{Sun}}, \bibnamefont{and}
  \bibinfo{author}{\bibfnamefont{R.~L.} \bibnamefont{Cone}},
  \bibinfo{note}{unpublished.}

\bibitem[{\citenamefont{{Wang} et~al.}(1996)\citenamefont{{Wang}, {Ruan},
  {Tsuboi}, and {Ter-Mikirtychev}}}]{wang96}
\bibinfo{author}{\bibfnamefont{X.}~\bibnamefont{{Wang}}},
  \bibinfo{author}{\bibfnamefont{Y.}~\bibnamefont{{Ruan}}},
  \bibinfo{author}{\bibfnamefont{T.}~\bibnamefont{{Tsuboi}}}, \bibnamefont{and}
  \bibinfo{author}{\bibfnamefont{V.~V.} \bibnamefont{{Ter-Mikirtychev}}}, in
  \emph{\bibinfo{booktitle}{Society of Photo-Optical Instrumentation Engineers
  (SPIE) Conference Series}}, edited by
  \bibinfo{editor}{\bibfnamefont{M.}~\bibnamefont{{Eich}}},
  \bibinfo{editor}{\bibfnamefont{B.~H.} \bibnamefont{{Chai}}},
  \bibnamefont{and} \bibinfo{editor}{\bibfnamefont{M.}~\bibnamefont{{Jiang}}}
  (\bibinfo{year}{1996}), vol. \bibinfo{volume}{2897} of
  \emph{\bibinfo{series}{Presented at the Society of Photo-Optical
  Instrumentation Engineers (SPIE) Conference}}, pp. \bibinfo{pages}{226--230}.

\bibitem[{\citenamefont{K{\"o}nz et~al.}(2003)\citenamefont{K{\"o}nz, Sun,
  Thiel, Cone, Equall, Hutcheson, and Macfarlane}}]{konz03}
\bibinfo{author}{\bibfnamefont{F.}~\bibnamefont{K{\"o}nz}},
  \bibinfo{author}{\bibfnamefont{Y.}~\bibnamefont{Sun}},
  \bibinfo{author}{\bibfnamefont{C.~W.} \bibnamefont{Thiel}},
  \bibinfo{author}{\bibfnamefont{R.~L.} \bibnamefont{Cone}},
  \bibinfo{author}{\bibfnamefont{R.~W.} \bibnamefont{Equall}},
  \bibinfo{author}{\bibfnamefont{R.~L.} \bibnamefont{Hutcheson}},
  \bibnamefont{and} \bibinfo{author}{\bibfnamefont{R.~M.}
  \bibnamefont{Macfarlane}}, \bibinfo{journal}{Physical Review B}
  \textbf{\bibinfo{volume}{68}}, \bibinfo{pages}{085109}
  (\bibinfo{year}{2003}).

\bibitem[{\citenamefont{Longdell}(2003)}]{longdell03}
\bibinfo{author}{\bibfnamefont{J.~J.} \bibnamefont{Longdell}}, Ph.D. thesis,
  \bibinfo{school}{The Australian National University} (\bibinfo{year}{2003}).

\bibitem[{\citenamefont{Wrigge et~al.}(2008)\citenamefont{Wrigge, Gerhardt,
  Hwang, Zumofen, and Sandoghdar}}]{wrigge07}
\bibinfo{author}{\bibfnamefont{G.}~\bibnamefont{Wrigge}},
  \bibinfo{author}{\bibfnamefont{I.}~\bibnamefont{Gerhardt}},
  \bibinfo{author}{\bibfnamefont{J.}~\bibnamefont{Hwang}},
  \bibinfo{author}{\bibfnamefont{G.}~\bibnamefont{Zumofen}}, \bibnamefont{and}
  \bibinfo{author}{\bibfnamefont{V.}~\bibnamefont{Sandoghdar}},
  \bibinfo{journal}{Nature Physics 4, 60 - 66}  (\bibinfo{year}{2008}).

\bibitem[{\citenamefont{Alnis et~al.}(2008)\citenamefont{Alnis, Matveev,
  Kolachevsky, Udem, and H{\"a}nsch}}]{alnis08}
\bibinfo{author}{\bibfnamefont{J.}~\bibnamefont{Alnis}},
  \bibinfo{author}{\bibfnamefont{A.}~\bibnamefont{Matveev}},
  \bibinfo{author}{\bibfnamefont{N.}~\bibnamefont{Kolachevsky}},
  \bibinfo{author}{\bibfnamefont{T.}~\bibnamefont{Udem}}, \bibnamefont{and}
  \bibinfo{author}{\bibfnamefont{T.~W.} \bibnamefont{H{\"a}nsch}},
  \bibinfo{journal}{Phys. Rev. A} \textbf{\bibinfo{volume}{77}},
  \bibinfo{pages}{053809} (\bibinfo{year}{2008}).

\bibitem[{\citenamefont{{Bjorklund}}(1980)}]{bjorklund80}
\bibinfo{author}{\bibfnamefont{G.~C.} \bibnamefont{{Bjorklund}}},
  \bibinfo{journal}{Optics Letters} \textbf{\bibinfo{volume}{5}},
  \bibinfo{pages}{15} (\bibinfo{year}{1980}).




\end{thebibliography}

\end{document}